\documentstyle[12pt]{article}
\begin{document}
\def\beq{\begin{equation}}
\def\eeq{\end{equation}}
\def\bea{\begin{eqnarray}}
\def\eea{\end{eqnarray}}
\def\ve{\vert}
\def\vel{\left|}
\def\ver{\right|}
\def\nnb{\nonumber}
\def\ga{\left(}
\def\dr{\right)}
\def\aga{\left\{}
\def\adr{\right\}}
\def\rar{\rightarrow}
\def\nnb{\nonumber}
\def\la{\langle}
\def\ra{\rangle}
\def\lla{\left<}
\def\rra{\right>}
\def\ba{\begin{array}}
\def\ea{\end{array}}
\def\tep{$B \rar K \ell^+ \ell^-$}
\def\tepm{$B \rar K \mu^+ \mu^-$}
\def\tept{$B \rar K \tau^+ \tau^-$}
\def\ds{\displaystyle}



\newskip\humongous \humongous=0pt plus 1000pt minus 1000pt
\def\caja{\mathsurround=0pt}
\def\eqalign#1{\,\vcenter{\openup1\jot
\caja   \ialign{\strut \hfil$\displaystyle{##}$&$
\displaystyle{{}##}$\hfil\crcr#1\crcr}}\,}


\def\simlt{\stackrel{<}{{}_\sim}}
\def\simgt{\stackrel{>}{{}_\sim}}



\def\bos{\lower 0.5cm\hbox{{\vrule width 0pt height 1.2cm}}}
\def\boss{\lower 0.35cm\hbox{{\vrule width 0pt height 1.cm}}}
\def\aaa{\lower 0.cm\hbox{{\vrule width 0pt height .7cm}}}
\def\dol{\lower 0.4cm\hbox{{\vrule width 0pt height .5cm}}}



\renewcommand{\textfraction}{0.2}    
\renewcommand{\topfraction}{0.8}   
\renewcommand{\bottomfraction}{0.4}   
\renewcommand{\floatpagefraction}{0.8}
\newcommand\mysection{\setcounter{equation}{0}\section}

\def\baeq{\begin{appeq}}     \def\eaeq{\end{appeq}}  
\def\baeeq{\begin{appeeq}}   \def\eaeeq{\end{appeeq}}
\newenvironment{appeq}{\beq}{\eeq}   
\newenvironment{appeeq}{\beeq}{\eeeq}
\def\bAPP#1#2{
 \markright{APPENDIX #1}
 \addcontentsline{toc}{section}{Appendix #1: #2}
 \medskip
 \medskip
 \begin{center}      {\bf\LARGE Appendix  :}{\quad\Large\bf #2}
\end{center}
 \renewcommand{\thesection}{#1.\arabic{section}}
\setcounter{equation}{0}
        \renewcommand{\thehran}{#1.\arabic{hran}}
\renewenvironment{appeq}
  {  \renewcommand{\theequation}{#1.\arabic{equation}}
     \beq
  }{\eeq}
\renewenvironment{appeeq}
  {  \renewcommand{\theequation}{#1.\arabic{equation}}
     \beeq
  }{\eeeq}
\nopagebreak \noindent}

\def\eAPP{\renewcommand{\thehran}{\thesection.\arabic{hran}}}

\renewcommand{\theequation}{\arabic{equation}}
\newcounter{hran}
\renewcommand{\thehran}{\thesection.\arabic{hran}}

\def\bmini{\setcounter{hran}{\value{equation}}
\refstepcounter{hran}\setcounter{equation}{0}
\renewcommand{\theequation}{\thehran\alph{equation}}\begin{eqnarray}}
\def\bminiG#1{\setcounter{hran}{\value{equation}}
\refstepcounter{hran}\setcounter{equation}{-1}
\renewcommand{\theequation}{\thehran\alph{equation}}
\refstepcounter{equation}\label{#1}\begin{eqnarray}}


\newskip\humongous \humongous=0pt plus 1000pt minus 1000pt
\def\caja{\mathsurround=0pt}
\def\eqalign#1{\,\vcenter{\openup1\jot
\caja   \ialign{\strut \hfil$\displaystyle{##}$&$
\displaystyle{{}##}$\hfil\crcr#1\crcr}}\,}


\title{ {\Large {\bf 
General analysis of time evolution of decay spectrum in
$B^0 \rar \pi^+ \pi^- \ell^+ \ell^-$} } }

\author{\vspace{1cm}\\
{\small T. M. Aliev \thanks
{e-mail: taliev@metu.edu.tr}\,\,,
A. \"{O}zpineci \thanks
{e-mail: ozpineci@newton.physics.metu.edu.tr}\,\,,
M. Savc{\i} \thanks
{e-mail: savci@metu.edu.tr}} \\
{\small Physics Department, Middle East Technical University} \\
{\small 06531 Ankara, Turkey} }
\date{}

\begin{titlepage}
\maketitle
\thispagestyle{empty}

\begin{abstract}
Using the general, model independent form of effective Hamiltonian,
the decay spectrum of $B^0 \rar \pi^+ \pi^- \ell^+ \ell^-$
decay is studied. The sensitivity of experimentally measurable asymmetries
to the new Wilson coefficients and time is studied. It is observed that
different asymmetries are sensitive to the new Wilson coefficients and they
can serve as an efficient tool for establishing new physics beyond SM.
\end{abstract}

~~~PACS number(s): 13.20.He, 12.60.--i
\end{titlepage}

\section{Introduction}
CP violation has been observed in the neutral kaon systems \cite{R1,R2}. Great
effort is devoted to the study of possible signals of CP violation in $B$
system, which will provide invaluable information about the origin of CP
violation and is one of the most promising research area 
having the potential of establishing new physics beyond the standard model
(SM). Started operating, two $B$--factories BaBar and Belle
open an excited era for a comprehensive study of $B$ meson physics 
and especially its rare decays.  The main
physics program of these factories constitutes a detailed study of CP
violation in $B_d$ meson and precise measurement of rare
flavor--changing neutral current (FCNC) processes. Furthermore, the ultimate
goal of these studies is to look for the inconsistencies within the SM
(see for example \cite{R3}), in particular, to find indications for indications for new
physics in the flavor and CP violating sectors. The factories mentioned
above have both already signaled the first evidences about the CP
violation in neutral $B$ meson decays \cite{R4}.  

One efficient way for detecting CP violation in the neutral $B$ meson decays
is the well known Dalitz plot asymmetry \cite{R5}. The main advantage of the
Dalitz plot asymmetries from the partial rate asymmetries is that they might
be present even when partial rate asymmetries vanish.
It should be noted that such CP--violating asymmetries in the angular
variable $\varphi$ ($\varphi$ is the angle between the $\ell^+ \ell^-$ and $\pi^+ \pi^-$
planes) in the $K \rar \pi^+ \pi^- \ell^+ \ell^-$ decay was measured
\cite{R2}. In a recent report, NA48 Collaboration \cite{R6} confirms large 
CP--violating
effect in the $K \rar \pi^+ \pi^- \ell^+ \ell^-$ decay. Time evolution of
the decay spectrum in $K^0~(\bar K^0 \rar \pi^+ \pi^- \ell^+ \ell^-)$ was
investigated in framework of the SM in \cite{R7}. 

In this connection several interesting questions come into mind immediately: how
does the angular variable asymmetry ${\cal A}_\varphi$ evolve with time for the
neutral $B$ meson decay? How do new
physics effects change the Dalitz plot asymmetry? Can we measure this
asymmetry in $B$--factories? This paper is devoted to answering these
questions by analyzing the time dependence of the decays $B^0,~\bar B^0 \rar
\pi^+ \pi^- \ell^+ \ell^-$ in detail. It should be noted that the $B \rar
\pi^+ \pi^- \ell^+ \ell^-$ decay in SM and beyond were studied in \cite{R8}
and \cite{R9}, respectively. 

The paper is organized as follows. In section 2 we present the general
expression of the Dalitz plot asymmetries for the $B^0~(\bar B^0 \rar\pi^+
\pi^- \ell^+ \ell^-)$ decays using the most general, model
independent form of the effective Hamiltonian. In section 3 we study the
sensitivity of the Dalitz plot asymmetries on the new Wilson coefficients.

\section{Time evolution of the decay spectrum of 
$B^0 \rar\pi^+\pi^- \ell^+ \ell^-$}
The matrix element for the $B \rar \rho(\rar \pi^+ \pi^-) \ell^+ \ell^-$
decay is described by the $b \rar d \ell^+ \ell^-$ transition at quark
level. Following \cite{R9}--\cite{R13}, the effective Hamiltonian of the $b
\rar d \ell^+ \ell^-$ transition can be written as the sum of the SM and new
physics contribution, in a general model independent way in the following
form
\bea
\label{matel}
{\cal H}_{eff} &=& \frac{G\alpha}{\sqrt{2} \pi}
 V_{td}V_{tb}^\ast
\Bigg\{ C_{SL} \, \bar d i \sigma_{\mu\nu} \frac{q^\nu}{q^2}\, L \,b
\, \bar \ell \gamma^\mu \ell + C_{BR}\, \bar d i \sigma_{\mu\nu}
\frac{q^\nu}{q^2} \,R\, b \, \bar \ell \gamma^\mu \ell \nnb \\
&&+C_{LL}^{tot}\, \bar d_L \gamma_\mu b_L \,\bar \ell_L \gamma^\mu \ell_L +
C_{LR}^{tot} \,\bar d_L \gamma_\mu b_L \, \bar \ell_R \gamma^\mu \ell_R +
C_{RL} \,\bar d_R \gamma_\mu b_R \,\bar \ell_L \gamma^\mu \ell_L \nnb \\
&&+C_{RR} \,\bar d_R \gamma_\mu b_R \, \bar \ell_R \gamma^\mu \ell_R +
C_{LRLR} \, \bar d_L b_R \,\bar \ell_L \ell_R +
C_{RLLR} \,\bar d_R b_L \,\bar \ell_L \ell_R \\
&&+C_{LRRL} \,\bar d_L b_R \,\bar \ell_R \ell_L +
C_{RLRL} \,\bar d_R b_L \,\bar \ell_R \ell_L+
C_T\, \bar d \sigma_{\mu\nu} b \,\bar \ell \sigma^{\mu\nu}\ell \nnb \\
&&+i C_{TE}\,\epsilon^{\mu\nu\alpha\beta} \bar d \sigma_{\mu\nu} b \,
\bar \ell \sigma_{\alpha\beta} \ell  \Bigg\}~, \nnb
\eea
where the chiral projection operators $L$ and $R$ in (\ref{matel}) are
defined as
\bea
L = \frac{1-\gamma_5}{2} ~,~~~~~~ R = \frac{1+\gamma_5}{2}\nnb~,
\eea
and $C_X$ are the coefficients of the four--Fermi interactions. 
Here $q=p_+ + p_-$ and $p_+$ and $p_-$ are the
four momenta of $\ell^+$ and $\ell^-$, respectively.
The first two of
these coefficients, $C_{SL}$ and $C_{BR}$, are the nonlocal Fermi
interactions which correspond to $-2 m_s C_7^{eff}$ and $-2 m_b C_7^{eff}$
in the SM, respectively. The next four terms
with coefficients $C_{LL}$, $C_{LR}$, $C_{RL}$ and $C_{RR}$ describe the
vector type interactions. Two of these
vector interactions containing $C_{LL}^{tot}$ and $C_{LR}^{tot}$ do already
exist in the SM
in combinations of the form $(C_9^{eff}-C_{10})$ and $(C_9^{eff}+C_{10})$.
Therefore, $C_{LL}^{tot}$ and $C_{LR}^{tot}$ describe the
sum of the contributions from SM and the new physics and they are defined as
\bea
C_{LL}^{tot} &=& C_9^{eff} - C_{10} + C_{LL}~, \nnb \\
C_{LR}^{tot} &=& C_9^{eff} + C_{10} + C_{LR}~.
\eea
The terms with
coefficients $C_{LRLR}$, $C_{RLLR}$, $C_{LRRL}$ and $C_{RLRL}$ describe
the scalar type interactions. The remaining last two terms leaded by the
coefficients $C_T$ and $C_{TE}$, obviously, describe the tensor type
interactions. 

The matrix element for the the process $B \rar \pi^+
\pi^- \ell^+ \ell^-$ can be obtained from the matrix element $B \rar \rho
\ell^+ \ell^-$, by the subsequent decay of the $\rho$ meson to pion pair
whose interaction Hamiltonian is
\bea
\label{h1}
{\cal H} = g_{\rho\pi\pi} (Q \varepsilon)~,
\eea
where $\varepsilon^\mu$ is the the polarization four vector of the $\rho$
meson, $Q=p_{\pi^+}-p_{\pi^-}$ and $p_{\pi^+}$ and $p_{\pi^-}$ are the
four momenta of $\pi^+$ and $\pi^-$, respectively.  
Hence, in calculating the 4--body decay amplitude for the $B \rar \pi^+
\pi^- \ell^+ \ell^-$ process, the matrix element of the $B \rar \rho \ell^+
\ell^-$ decay is needed. For this purpose it is enough to sandwich the
effective Hamiltonian (\ref{matel}) between the initial and final meson
states. In other words, the following matrix elements
\bea
\label{roll}
&&\lla \rho \vel \bar d \gamma_\mu (1 \pm \gamma_5)
b \ver B \rra~,\nnb \\
&&\lla \rho \vel \bar d
(1 \pm \gamma_5) b \ver B \rra~, \nnb \\
&&\lla \rho \vel \bar d \sigma_{\mu\nu}(1 \pm \gamma_5) b
\ver B \rra~, \nnb
\eea
need to be calculated in order to calculate the decay amplitude for the $B
\rar \rho \ell^+ \ell^-$ process. These matrix elements can be written in
terms of the form factors as 
\bea
\lefteqn{
\label{ilk}
\lla \rho(P,\varepsilon) \vel \bar d \gamma_\mu
(1 \pm \gamma_5) b \ver B(p_B) \rra =} \nnb \\
&&- \epsilon_{\mu\nu\lambda\sigma} \varepsilon^{\ast\nu} P^\lambda
q^\sigma
\frac{2 V(q^2)}{m_B+m_{\rho}} \pm i \varepsilon_\mu^\ast (m_B+m_{\rho})
A_1(q^2) \\
&&\mp i (p_B + P)_\mu (\varepsilon^\ast q)
\frac{A_2(q^2)}{m_B+m_{\rho}}
\mp i q_\mu \frac{2 m_{\rho}}{q^2} (\varepsilon^\ast q)
\left[A_3(q^2)-A_0(q^2)\right]~.\nnb
\eea
Multiplying both sides of Eq. (\ref{ilk}) with $q_\mu$ and using equation of
motion, we get 
\bea
\lefteqn{
\label{iki}
\lla \rho(P,\varepsilon) \vel \bar d
(1 \pm \gamma_5) b \ver B(p_B) \rra =}\\
&&\frac{i}{m_b} ( \varepsilon^{\ast} q) \Bigg\{ \mp (m_B+m_{\rho}) A_1(q^2) 
\pm (m_B^2-P^2) \frac{A_2(q^2)}{m_B+m_{\rho}}
\pm 2 m_\rho \Big[A_3(q^2) - A_0(q^2) \Big] \Bigg\}~.\nnb
\eea
The remaining  matrix element is defined as
\bea
\lefteqn{
\label{ucc}
\lla \rho(P,\varepsilon) \vel \bar d \sigma_{\mu\nu}
 b \ver B(p_B) \rra =} \nnb \\
&&i \epsilon_{\mu\nu\lambda\sigma}  \Bigg\{ - 2 T_1(q^2)
{\varepsilon^\ast}^\lambda (p_B + P)^\sigma +
\frac{2}{q^2} (m_B^2-P^2) \Big[ T_1(q^2) - T_2(q^2) \Big]
{\varepsilon^\ast}^\lambda
q^\sigma \\
&&- \frac{4}{q^2} \Bigg[ T_1(q^2) - T_2(q^2) -
\frac{q^2}{m_B^2-P^2}
T_3(q^2) \Bigg] (\varepsilon^\ast q) P^\lambda q^\sigma \Bigg\}~.
\nnb
\eea
Moreover, the matrix element $\lla \rho(P,\varepsilon) \vel \bar d
\sigma_{\mu\nu} \gamma_5 b \ver B(p_B) \rra$ can be calculated from Eq.
(\ref{ucc}) using the identity
\bea
\sigma_{\mu\nu} \gamma_5 = -\frac{i}{2} \epsilon_{\mu\nu\alpha\beta}
\sigma^{\alpha\beta} ~,\nnb
\eea
which yields
\bea
\lefteqn{
\label{dort}
\lla \rho(P,\varepsilon) \vel \bar d \sigma_{\mu\nu}\gamma_5
 b \ver B(p_B) \rra =} \nnb \\
&& 2 \Bigg\{  T_1(q^2)\Big[
\varepsilon^\ast_\mu (p_B + P)_\nu - 
\varepsilon^\ast_\nu (p_B + P)_\mu \Big]
- \frac{1}{q^2} (m_B^2-P^2) (\varepsilon^\ast_\mu q_\nu -
\varepsilon^\ast_\nu q_\mu) \Big[ T_1(q^2) - T_2(q^2) \Big] \nnb \\
&&+ \frac{2}{q^2} \Bigg[ T_1(q^2) - T_2(q^2) -
\frac{q^2}{m_B^2-P^2}
T_3(q^2) \Bigg] (\varepsilon^\ast q) (P^\mu q^\nu -P^\nu q^\mu)  \Bigg\}~.
\eea
Using Eqs. (\ref{h1}--\ref{dort}) for the matrix element
$B \rar \pi^+ \pi^- \ell^+ \ell^-$ decay we get 
\bea
\lefteqn{
\label{had}
{\cal M}(B\rightarrow \pi^+ \pi^- \ell^+ \ell^-) =
\frac{G \alpha}{4 \sqrt{2}} \, V_{tb} V_{td}^\ast \,
\ds\frac{g^{\nu\alpha} - P^\nu P^\alpha/m_\rho^2}
{P^2-m_\rho^2+ i m_\rho \Gamma}\,g_{\rho\pi\pi} Q_\alpha } \nnb \\
&&\times \Bigg\{
\bar \ell \gamma^\mu (1-\gamma_5) \ell \, \Big[
-2 {\cal V}_{L_1} \epsilon_{\mu\nu\lambda\sigma}
P^\lambda q^\sigma
- i {\cal V}_{L_2} g_{\mu\nu} +
i {\cal V}_{L_3} (2 P + q)_\mu q_\nu + i {\cal V}_{L_4} q_\mu q_\nu
\Big] \nnb \\
&&+ \bar \ell \gamma^\mu (1+\gamma_5) \ell \, \Big[
-2 {\cal V}_{R_1} \epsilon_{\mu\nu\lambda\sigma}                      
P^\lambda q^\sigma
- i {\cal V}_{R_2} g_{\mu\nu} +
i {\cal V}_{R_3} (2 P + q)_\mu q_\nu + i {\cal V}_{R_4} q_\mu q_\nu
\Big] \nnb \\
&&+\bar \ell (1-\gamma_5) \ell \, ( i S_L) q_\nu
+ \bar \ell (1+\gamma_5) \ell \, q_\nu ( i S_R)\nnb \\
&&+4 \bar \ell \sigma_{\mu\beta}  \ell \Big( i C_T
\epsilon^{\mu\beta\lambda\sigma}
\Big) \Big[ -2 T_1 g_{\lambda\nu} (2 P +q)_\sigma +
B_6 g_{\lambda\nu} q_\sigma -
B_7 P_\lambda q_\sigma q_\nu\Big] \nnb \\
&&+16 C_{TE} \bar \ell \sigma^{\mu\beta}  \ell \Big[ -2 T_1
g_{\mu\nu} (2 P+q)_\beta  +B_6 g_{\mu\nu} q_\beta -
B_7 q_\nu P_\mu q_\beta \Big]
\Bigg\}~,
\eea
where ${\cal{V}}_{L_i}$ and ${\cal{V}}_{R_i}$ are the coefficients of left--
and right--handed
leptonic currents with vector structure, respectively, $S_{L,R}$ are the
coefficients of
the leptonic scalar currents and the last two terms correspond to the
contribution of the leptonic tensor currents. These new coefficients are
functions of the Wilson coefficients and the form factors introduced in
defining the hadronic matrix elements above. Their explicit expressions are
given by
\bea
{\cal{V}}_{L_1} &=& (C_{LL}^{tot} + C_{RL}) \frac{V}{m_B+m_{\rho}} -
2 (C_{BR} + C_{SL}) \frac{T_1}{q^2}  , \nnb \\
{\cal{V}}_{L_2} &=& (C_{LL}^{tot} - C_{RL}) (m_B+m_{\rho}) A_1 - 2 (C_{BR} -
C_{SL})
\frac{T_2}{q^2} (m_B^2-m_{\rho}^2) ~, \nnb \\
{\cal{V}}_{L_3} &=& \frac{C_{LL}^{tot} - C_{RL}}{m_B+m_{\rho}} A_2 - 2 (C_{BR} -
C_{SL})
\frac{1}{q^2}  \left[ T_2 + \frac{q^2}{m_B^2-m_{\rho}^2} T_3 \right]~,
\nnb \\
{\cal{V}}_{L_4} &=& (C_{LL}^{tot} - C_{RL}) \frac{2 m_{\rho}}{q^2} (A_3-A_0)-
2 (C_{BR} - C_{SL}) \frac{T_3}{q^2}~, \nnb \\
{\cal{V}}_{R_1} &=& {\cal{V}}_{L_1} ( C_{LL}^{tot} \rar C_{LR}^{tot}~,~~C_{RL} \rar
C_{RR})~, \nnb \\
{\cal{V}}_{R_2} &=& {\cal{V}}_{L_2} ( C_{LL}^{tot} \rar C_{LR}^{tot}~,~~C_{RL} \rar
C_{RR})~,\nnb \\
{\cal{V}}_{R_3} &=& {\cal{V}}_{L_3} ( C_{LL}^{tot} \rar C_{LR}^{tot}~,~~C_{RL} \rar
C_{RR})~,\nnb \\
{\cal{V}}_{R_4} &=& {\cal{V}}_{L_4} ( C_{LL}^{tot} \rar C_{LR}^{tot}~,~~C_{RL} \rar
C_{RR})~,\nnb \\
{\cal{S}}_{L} &=& - ( C_{LRRL} - C_{RLRL}) \, \frac{1}{m_b} \,
\Bigg[ (m_B + m_{\rho}) A_1 - (m_B^2-P^2) \frac{A_2}{m_B + m_{\rho}}-
2 m_{\rho}(A_3- A_0) \Bigg] ~,\nnb \\
{\cal{S}}_{R} &=& - ( C_{LRLR} - C_{RLLR}) \, \frac{1}{m_b} \,
\Bigg[ (m_B + m_{\rho}) A_1 - (m_B^2-P^2) \frac{A_2}{m_B + m_{\rho}}-
2 m_{\rho}(A_3- A_0) \Bigg] ~,\nnb
\eea
where dependence on $q^{2}$ is implied. 
To obtain the amplitudes for the $\bar B^0 \rar \pi^+ \pi^- \ell^+ \ell^-$
decay, it is enough to use CPT invariance and make the replacements in Eq.
(\ref{had})
\bea
\label{trn}
{\cal{V}}_{L_1;R_1} &\rar& - \overline {\cal{V}}_{L_1;R_1}~,\nnb \\
{\cal{V}}_{L_i;R_i} &\rar& \overline{\cal{V}}_{L_i;R_i}~,~~~(i=2,3,4)~,\nnb \\
{\cal{S}}_{L;R} &\rar& \overline {\cal{S}}_{R;L}~,\\
B_{6;7;8}  &\rar& \bar B_{6;7;8}~,\nnb \\
B_{9;10;11}  &\rar& - \bar B_{9;10;11}~,\nnb
\eea
where the meaning of the bar can be explained as follows: If any arbitrary
one of the coefficients ${\cal{V}}_{L_i;R_i}$ is represented in the form
\bea 
{\cal{V}}_{L_i;R_i} = \vel {\cal{V}}_{L_i;R_i} \ver
e^{i \delta_{L_i;R_i}} e^{i \phi_{L_i;R_i}}~, \nnb
\eea
where $\delta$ and $\phi$ are the strong and the weak phases, respectively,
then
\bea
\overline {\cal{V}}_{L_i;R_i} = \vel {\cal{V}}_{L_i;R_i} \ver
e^{i \delta_{L_i;R_i}} e^{- i \phi_{L_i;R_i}}~. \nnb
\eea
The starting point in analysis of the time evolution spectrum of the
$B^0 (\bar B^0) \rar \pi^+ \pi^- \ell^+ \ell^-$ decay is Eq.
(\ref{had}). The time evolution of the $B$ mesons is given by
\bea
\vel B^0(t) \rra &=& g_+(t) \vel B^0 \rra + 
\frac{1}{\xi} g_-(t) \vel \bar B^0 \rra  ~, \nnb \\
\vel \bar B^0(t) \rra &=& g_+(t) \vel \bar B^0 \rra + 
\xi g_-(t) \vel B^0 \rra  ~, \nnb
\eea
with 
\bea
g_\pm = exp \left[ - \ga \frac{\Gamma_B}{2} - i m_B \dr t \right] 
\left\{\cos \ga \frac{\Delta m}{2} t \dr ~ ;~
      \sin \ga \frac{\Delta m}{2} t \dr \right\}~,\nnb
\eea
and $\xi=p/q$. Here $m_B(\Gamma_B)$ and $\Delta m_B(\Delta \Gamma_B)$ are
the average and the difference of the masses (widths) of the two mass
eigenstates $B_H$ and $B_L$, respectively.

The matrix element of the time dependent $B^0(t)  \rar \pi^+ \pi^- \ell^+
\ell^-$ decay can be obtained from Eq. (\ref{had}) by making the following
replacements. 
\bea
{\cal{V}}_{L_1;R_1} \rar {\cal{V}}_{L_1;R_1}(t) &=& g_+(t)\, {\cal{V}}_{L_1;R_1}
-  \frac{1}{\xi} \,g_-(t) \, \overline {\cal{V}}_{L_1;R_1}~, \nnb \\ 
{\cal{V}}_{L_i;R_i} \rar {\cal{V}}_{L_i;R_i}(t) &=&g_+(t) \, {\cal{V}}_{L_i;R_i}
+  \frac{1}{\xi} \, g_-(t) \,\overline {\cal{V}}_{L_i;R_i}~,~~~(i=2,3,4)~,\nnb \\
{\cal{S}}_{L;R} \rar {\cal{S}}_{L;R} (t) &=& g_+(t)\, {\cal{S}}_{L;R}   
+  \frac{1}{\xi} \,g_-(t) \, \overline {\cal{S}}_{R;L}~,\\
B_{6;7;8} \rar B_{6;7;8}(t) &=& g_+(t)\, B_{6;7;8} + 
\frac{1}{\xi} \, g_-(t) \,\bar B_{6;7;8}~, \nnb \\
B_{9;10;11} \rar B_{9;10;11}(t) &=&  g_+(t)\, B_{9;10;11} -
\frac{1}{\xi} \, g_-(t) \,\bar B_{9;10;11}~.\nnb
\eea
Similarly the matrix element for the $\bar B^0(t)  \rar \pi^+ \pi^- \ell^+
\ell^-$ decay can be derived from the respective matrix element
${\cal M} (\bar B^0 \rar \pi^+ \pi^- \ell^+\ell^-)$ with the help of the
replacements
\bea
\label{trnn} 
\overline {\cal{V}}_{L_1;R_1} \rar \overline {\cal{V}}_{L_1;R_1}(t) &=& g_+(t)\,
\overline {\cal{V}}_{L_1;R_1}            
- \xi \,g_-(t) \,{\cal{V}}_{L_1;R_1}~, \nnb \\
\overline {\cal{V}}_{L_i;R_i} \rar \overline {\cal{V}}_{L_i;R_i}(t) &=&g_+(t) \,
\overline {\cal{V}}_{L_i;R_i}    
+ \xi \, g_-(t) \,
{\cal{V}}_{L_i;R_i}~,~~~(i=2,3,4)~,\nnb \\
\overline {\cal{S}}_{L;R} \rar \overline {\cal{S}}_{L;R} (t) &=& g_+(t)\, 
\overline {\cal{S}}_{L;R} 
+ \xi \,g_-(t) \,{\cal{S}}_{R;L}~,\\
\bar B_{6;7;8} \rar \bar B_{6;7;8}(t) &=& g_+(t)\, \bar B_{6;7;8} + 
\xi \, g_-(t) \,B_{6;7;8}~, \nnb \\
\bar B_{9;10;11} \rar \bar B_{9;10;11}(t) &=&  g_+(t)\, \bar B_{9;10;11} -
\xi \, g_-(t) \,B_{9;10;11}~.\nnb
\eea
Using the formalism for the $K_{\ell_4}$ decay \cite{R14}, from the matrix
element (\ref{had}) one can obtain the $B^0 \rar \pi^+ \pi^- \ell^+\ell^-$
decay rate in terms of the following five variables: invariant mass 
$s_M = P^2 = (p_{\pi^+}+p_{\pi^-})^2$ of $\pi^+ \pi^-$ pair, invariant mass 
$s_\ell  = q^2 = (p_+ + p_-)^2$ of $\ell^+\ell^-$ pair; the angle $\theta$
between $\vec{p}_{\pi^+}$ and $\vec{p}_+ + \vec{p}_-$, measured with respect to
the  center of mass of $\pi^+ \pi^-$ pair; the angle $\theta_\ell$ between
$\vec{p}_+$ and $\vec{p}_{\pi^+} + \vec{p}_{\pi^-}$ , measured with respect to   
the  center of mass of $\ell^+\ell^-$ pair and $\varphi$ is the angle between
the normals to the $\pi^+ \pi^-$ and $\ell^+\ell^-$ planes. The final
four--body phase volume in terms of the above--mentioned five variables can
be written as
\bea
\label{XPS}
d X_{PS} = \frac{1}{2 (4\pi)^6 m_B^2} X \beta \beta_\ell \,
ds_M \, ds_\ell \,d(\cos\theta)\, d(\cos\theta_\ell)\,d\varphi~,
\eea
where
\bea
X &=& \left[ \ga P q\dr^2 - s_M s_L \right]^{1/2}~, \nnb \\
\beta &=& \frac{1}{s_M} \, \lambda^{1/2}(s_M,m_\pi^2,m_\pi^2)~,\nnb \\
\beta_\ell &=& \frac{1}{s_\ell} \, \lambda^{1/2}(s_\ell,m_\ell^2,m_\ell^2)~.\nnb
\eea
The bounds of the integration variables are
\bea
4 m_\pi^2 \!\!\!&\le&\!\!\! s_M \le m_B^2~, \nnb \\
4 m_\ell^2 \!\!\!&\le&\!\!\! s_\ell \le \ga m_B - \sqrt{s_M} \dr^2~, \nnb \\
0 \!\!\!&\le&\!\!\! \theta, \theta_\ell \le \pi~, \nnb \\
0 \!\!\!&\le&\!\!\! \varphi \le 2 \pi~.\nnb
\eea   
Different scalar quantities appearing in expression of the $\vel {\cal M}
\ver^2$ can be expressed in terms of the above--mentioned five variables as
\bea
P q &=& \frac{1}{2} (m_B^2 - s_M - s_\ell)~,\nnb \\
Q N &=& \beta \beta_\ell \left( Pq \, \cos\theta \, \cos\theta_\ell 
-\sqrt{s_M s_\ell} \, \sin\theta \,\sin\theta_\ell \,\cos\varphi\right)~,\nnb \\
p_B  p_\pm &=&  \frac{1}{2}\ga s_M + 
P q \pm X\, \beta\,\cos\theta \dr~,\nnb \\
q N &=& 0~,\nnb \\
Q q &=& X\, \beta\,\cos\theta~,\nnb \\
Q P &=& 0~,\nnb \\
N P &=& X\, \beta_\ell \, \cos\theta_\ell~,\nnb \\
Q^2 &=& -s_M \beta^2~,\nnb \\
N^2 &=& -s_\ell \beta_\ell^2~,\nnb \\
\epsilon_{\mu\nu\lambda\sigma} Q^\mu P^\nu N^\lambda q^\sigma &=&
-\sqrt{s_M s_\ell} \,X \, \beta \,\beta_\ell \, \sin\theta \, \sin\theta_\ell
\, \sin\varphi~,\nnb
\eea
where
\bea
N=p_+ - p_-~.\nnb
\eea
The differential decay rate for the $B^0(t) \rar \pi^+ \pi^- \ell^+\ell^-$
is (in the expression below leptons are taken to be massless. However in
calculation of the decay spectrum for $B^0(t) \rar \pi^+ \pi^-
\tau^+\tau^-$, the $\tau$ lepton mass is taken into account. But this
expression is quite lengthy and for this reason we do not present it in
the text)
\bea
d\Gamma &=& \vel \frac{G\alpha}{4 \sqrt{2} \pi} V_{tb}V_{td}^\ast \ver^2\, 
\frac{g_{\rho\pi\pi}^2}{\ga s_M - m_\rho^2 \dr^2 + \Gamma^2_\rho m_\rho^2}\,
\frac{1}{2^{14} \pi^6 m_B^3} \, X\, \beta \nnb\\
&&\times ds_M\,ds_\ell\, d(\cos\theta) \, d(\cos\theta_\ell)\, d\varphi~
{\cal I}~,
\eea  
where
\bea
\label{I}
\lefteqn{
{\cal I} = {\cal I}_1 \, \cos^2\!\theta \,\cos^2\!\theta_\ell +
{\cal I}_2 \, \cos^2\!\theta \,\cos\theta_\ell +
{\cal I}_3 \, \cos^2\!\theta
+{\cal I}_4 \, \sin^2\!\theta \,\sin^2\!\theta_\ell \, \sin(2\varphi)} \nnb \\
&&+{\cal I}_5 \, \sin(2\theta) \,\sin(2\theta_\ell) \, \cos\varphi
+ {\cal I}_6 \, \sin(2\theta) \,\sin(2\theta_\ell) \, \sin\varphi + 
{\cal I}_7 \, \sin(2\theta) \,\sin\theta_\ell \, \sin\varphi  \\              
&&+{\cal I}_8 \, \sin(2\theta) \,\sin\theta_\ell \, \cos\varphi +
{\cal I}_9 \, \cos^2\!\theta_\ell +{\cal I}_{10} \, \cos\theta_\ell
+{\cal I}_{11} \, \sin^2\!\theta \,\sin^2\!\theta_\ell \,\cos^2\!\varphi+
{\cal I}_{12}~,\nnb
\eea
and
\newpage
\bea
{\cal I}_1 &=& \beta^2 \Big\{ \lambda^2 \Big[ \vel {\cal V}_{L_1} \ver^2
- \vel {\cal V}_{L_3} \ver^2 + \vel {\cal V}_{R_1} \ver^2
- \vel {\cal V}_{R_3} \ver^2 \Big]
- 2 \lambda (P q) \Big[ 2\, \mbox{\rm Re} \ga {\cal V}_{L_2}{\cal
V}_{L_3}^\ast\dr \nnb \\
&+& 2 \, \mbox{\rm Re} 
\ga {\cal V}_{R_2}{\cal V}_{R_3}^\ast\dr\Big]
- 4 \ga P q\dr^2 \Big[ \lambda \ga \vel {\cal V}_{L_1} \ver^2
+\vel {\cal V}_{R_1} \ver^2 \dr + \vel {\cal V}_{L_2} \ver^2 
+\vel {\cal V}_{R_2} \ver^2 \Big] \Big\}~, \nnb \\ \nnb \\
{\cal I}_2 &=& 2 \beta^2 \sqrt{\lambda} \Big[\lambda - 4 \ga P q\dr^2
\Big] \Big[ - \mbox{\rm Re} \ga {\cal V}_{L_1}{\cal
V}_{L_2}^\ast\dr + \mbox{\rm Re} \ga {\cal V}_{R_1}{\cal
V}_{R_2}^\ast\dr \Big]~, \nnb \\ \nnb \\
{\cal I}_3 &=& \beta^2 \lambda\Big\{ \lambda \ga \vel {\cal V}_{L_3} \ver^2
+ \vel {\cal V}_{R_3} \ver^2 \dr + \vel {\cal V}_{R_2} \ver^2  -
4 \ga P q \dr \Big[ \mbox{\rm Re} \ga {\cal V}_{L_2}
{\cal V}_{L_3}^\ast \dr + \mbox{\rm Re} \ga {\cal V}_{R_2} 
{\cal V}_{R_3}^\ast\dr \Big] \Big\}~, \nnb \\ \nnb \\ 
{\cal I}_4 &=& 4 \beta^2 \sqrt{\lambda}\Big[
\mbox{\rm Im} \ga {\cal V}_{L_2} {\cal V}_{L_1}^\ast \dr
+ \mbox{\rm Im} \ga {\cal V}_{R_2} {\cal V}_{L_1}^\ast \dr 
\Big] ~, \nnb \\ \nnb \\
{\cal I}_5 &=& - \beta^2 \sqrt{s_M s_\ell}\,
\Big\{\lambda \Big[ \mbox{\rm Re} \ga {\cal V}_{L_2}{\cal V}_{L_3}^\ast
\dr + \mbox{\rm Re} \ga {\cal V}_{R_2}{\cal V}_{R_3}^\ast\dr \Big] -
2 \ga P q\dr \ga \vel {\cal V}_{L_2} \ver^2 + \vel {\cal V}_{R_2}
\ver^2 \dr \Big\}~, \nnb \\ \nnb \\
{\cal I}_6 &=& \beta^2 \sqrt{\lambda}\, \Big\{\lambda
\Big[ \mbox{\rm Im} \ga {\cal V}_{L_3}{\cal V}_{L_1}^\ast   
\dr + \mbox{\rm Im} \ga {\cal V}_{R_3}{\cal V}_{R_1}^\ast\dr \Big]
+ 2 \ga P q \dr \Big[\mbox{\rm Im} \ga {\cal V}_{L_1}{\cal
V}_{L_2}^\ast\dr \nnb \\ 
&+& \mbox{\rm Im} \ga {\cal V}_{R_1}{\cal V}_{R_2}^\ast\dr
\Big]\Big\}~, \nnb \\ \nnb \\
{\cal I}_7 &=& 2 \beta^2 \lambda \,\sqrt{s_M s_\ell}\,
\Big[ \mbox{\rm Im} \ga {\cal V}_{L_3}{\cal V}_{L_2}^\ast   
\dr - \mbox{\rm Im} \ga {\cal V}_{R_3}{\cal V}_{R_2}^\ast\dr 
\Big]~, \nnb \\ \nnb \\
{\cal I}_8 &=& 2 \beta^2 \sqrt{\lambda s_M s_\ell}\, \Big\{\lambda
\Big[ \mbox{\rm Re} \ga {\cal V}_{L_1}{\cal V}_{L_3}^\ast        
\dr - \mbox{\rm Re} \ga {\cal V}_{R_1}{\cal V}_{R_3}^\ast\dr \Big]        
+ 2 \ga P q \dr \Big[-\mbox{\rm Re} \ga {\cal V}_{L_1}{\cal
V}_{L_2}^\ast\dr \nnb \\ 
&+& \mbox{\rm Re} \ga {\cal V}_{R_1}{\cal V}_{R_2}^\ast\dr
\Big]\Big\}~, \nnb \\ \nnb \\
{\cal I}_9 &=& 4 \beta^2 s_\ell s_M \lambda \,
\Big[ \vel {\cal V}_{L_1} \ver^2 + \vel {\cal V}_{R_1} \ver^2 
\Big]~, \nnb \\ \nnb \\
{\cal I}_{10} &=& 8 \beta^2 \sqrt{\lambda} \,
\Big[ -\mbox{\rm Re} \ga {\cal V}_{L_1}{\cal V}_{L_2}^\ast
\dr + \mbox{\rm Re} \ga {\cal V}_{R_1}{\cal V}_{R_2}^\ast\dr 
\Big]~, \nnb \\ \nnb \\
{\cal I}_{11} &=& 4 \beta^2 s_\ell s_M \,
\Big[ \lambda \ga \vel {\cal V}_{L_1} \ver^2 + \vel {\cal V}_{R_1} \ver^2 
\dr - \vel {\cal V}_{L_2} \ver^2 - \vel {\cal V}_{R_2} \ver^2 
\Big]~, \nnb \\ \nnb \\
{\cal I}_{12} &=& 4 \beta^2 s_\ell s_M \,        
\Big[ \vel {\cal V}_{L_2} \ver^2 + \vel {\cal V}_{R_2} \ver^2 
\Big]~. \nnb
\eea
Taking the narrow resonance limit of $\rho$ meson, i.e., using the equations
\bea
\Gamma = \frac{g_{\rho\pi\pi}^2 m_\rho}{48 \pi} 
\ga 1 - \frac{4 m_\pi^2}{m_\rho^2} \dr^{3/2}~,~~~\mbox{\rm and},\nnb \\
\lim_{\Gamma \rar 0} \,
\frac{\Gamma m_\rho}{\ga s_M-m_\rho^2 \dr^2 + m_\rho^2 \Gamma^2}
= \pi \delta\ga s_M - m_\rho^2 \dr~,\nnb
\eea                  
we perform integration over $s_M$ easily in Eq. (\ref{I}). After integrating
over $s_M$, $s_\ell$ and $\theta$, we get for the angular distribution of
the $B^0(t)  \rar \pi^+ \pi^- \ell^+ \ell^-$ decay 
\bea
\label{ang}
\lefteqn{
\frac{d\Gamma}{d(cos\theta_\ell)\, d\varphi} =
\vel \frac{G \alpha}{4\sqrt{2} \pi}\, V_{tb}V_{td}^\ast \ver^2
\,\frac{1}{2^{14} \pi^6 m_B^3} \, \frac{48 \pi^2}
{m_\rho^2 \ga 1 - 4 m_\pi^2 / m_\rho^2 \dr^{1/2}} } \nnb \\
&&\times \Bigg\{\ga \frac{2}{3} {\cal I}_1^\prime + 2 {\cal I}_9^\prime \dr
\cos^2\!\theta_\ell + \ga \frac{2}{3} {\cal I}_2^\prime + 
2 {\cal I}_{10}^\prime \dr \cos\theta_\ell 
+ \frac{4}{3} {\cal I}_4^\prime \sin^2\!\theta_\ell \, \sin(2\varphi)\nnb \\
&&+\frac{2}{3} {\cal I}_{11}^\prime \sin^2\!\theta_\ell 
\Big[1+\cos(2\varphi)\Big] + \ga \frac{2}{3} {\cal I}_3^\prime
+ 2 {\cal I}_{12}^\prime \dr \Bigg\}~,
\eea            
where we have introduced the notation ${\cal I}_i^\prime \equiv 
\int X {\cal I}_i \, ds_\ell$. CP violation manifests itself in the
coefficients $\sim {\cal I}_4^\prime$ and $\sim {\cal I}_{11}^\prime$.
The dominant CP violation term is contained in the coefficient
$(4/3){\cal I}_4^\prime$. After integrating over $\cos\theta_\ell$ we obtain
the $\varphi$--distribution
\bea
\label{phi}  
\frac{d\Gamma}{d\varphi} \sim \frac{2}{3} \ga \frac{2}{3}  
{\cal I}_1^\prime + 2  {\cal I}_9^\prime \dr +
\frac{16}{9} {\cal I}_4^\prime \sin(2\varphi) +
\frac{8}{9} {\cal I}_{11}^\prime \Big[1+\cos(2\varphi)\Big] +
2  \ga \frac{2}{3}{\cal I}_3^\prime + 2  {\cal I}_{12}^\prime \dr~. \nnb
\eea 
As has already been noted the term $\sim \sin(2\varphi)$ is CP-- and
T--odd. Hence this term can be isolated by considering the following
asymmetry in terms of the differential decay rates of the $B$ meson with
respect to $\varphi$
\bea
{\cal A}_\varphi(t) = 
 \frac{\ds \left\{\int_0^{\pi/2} \, - \, \int_{\pi/2}^\pi \,
+  \int_\pi^{3\pi/2} \,- \, \int_{3\pi/2}^{2\pi} \right\}
\ds\frac{d\Gamma}{d\varphi}\, d\varphi }
{\ds \left\{\int_0^{\pi/2} \, + \, \int_{\pi/2}^\pi \,
+  \int_\pi^{3\pi/2} \,+ \, \int_{3\pi/2}^{2\pi} \right\}
\ds\frac{d\Gamma}{d\varphi} \, d\varphi}~.\nnb
\eea
The main feature of such CP violating asymmetries is that they can be
obtained by considering the sum of the differential decay rates of the $B$
and $\bar B$ rather than the difference of these rates, as usual.

Furthermore we study the dependence of the asymmetry between the spectrum
integrated rates of $B^0(t)  \rar \pi^+ \pi^- \ell^+ \ell^-$ and 
$\bar B^0(t)  \rar \pi^+ \pi^- \ell^+ \ell^-$ decays, on the new Wilson
coefficients, which is defined as
\bea
\label{A1}
{\cal A}_1 (t) = \frac{\ds 
\Gamma(B^0(t)  \rar \pi^+ \pi^- \ell^+ \ell^-) -
\Gamma (\bar B^0(t)  \rar \pi^+ \pi^- \ell^+ \ell^-)}
{\ds \Gamma (B^0(t)  \rar \pi^+ \pi^- \ell^+ \ell^-) +
\Gamma (\bar B^0(t)  \rar \pi^+ \pi^- \ell^+ \ell^-)}~.
\eea 

It should noted that it is also possible to construct a differential
asymmetry that isolates another combination of the terms containing
imaginary parts 
(see for example \cite{R15,R16}). Such terms, for example, 
can be isolated by considering the
difference distribution of the same hemisphere and opposite
hemisphere events. The corresponding asymmetry can be defined then as follows
\bea
\label{A2}
{\cal A}_2 (t) &=& \frac{1}{\Gamma}\,
\left\{ \int_0^\pi\, - \, \int_0^{2\pi} \right\} d\varphi \,
\left\{ \int_{-1}^0\, - \, \int_0^{+1} \right\} d(\cos\theta_\ell) \,
\left\{ \int_{-1}^0\, - \, \int_0^{+1} \right\} d(\cos\theta)\nnb \\ 
&\times& 
\frac{d\Gamma}{d\varphi\, d(\cos\theta_\ell) \,d(\cos\theta)}~.
\eea

\section{Numerical analysis}
In this section we study the dependence of the asymmetries ${\cal
A}_\varphi,~{\cal A}_1$ and ${\cal A}_2$ on new Wilson coefficients and time
for the $B^0(t) \rar \pi^+ \pi^- \ell^+ \ell^-$ decay. Before
presenting the numerical results we want to note that, although the
expressions for the asymmetries ${\cal A}_\varphi,~{\cal A}_1$ and ${\cal
A}_2$ are given for the most general case, i.e., each one of new Wilson 
coefficients might have new strong and weak phases. In our numerical
analysis we choose all new Wilson coefficients to be real and they all vary
in the region between $-4$ and $+4$. 

For the form factors which are needed in the course of performing  numerical
calculations, we have used the prediction of the light cone QCD sum rules.
In our numerical analysis 
we will use the results of the work \cite{R17} (see also \cite{R18,R19}) 
in which the form
factors are described by a three--parameter fit where the radiative
corrections up to leading twist contribution and
SU(3)--breaking effects are taken into account.
The $q^{2}$--dependence of the form factors which appear in our analysis
could
be parametrized as
\bea
\label{formfac}
F(s) = \frac{F(0)}{1-a_F\,s + b_F\, s^{2}}~, \nnb
\eea
where $s = q^2/m_B^2$ is the dilepton invariant mass in units of
$B$ meson mass, and the parameters $F(0)$, $a_F$ and $b_F$ are
listed in Table 1 for each
form factor.
\begin{table}[h]
\renewcommand{\arraystretch}{1.5}
\addtolength{\arraycolsep}{3pt}
$$  
\begin{array}{|l|ccc|}
\hline
& F(0) & a_F & b_F \\ \hline
A_0^{B \rar \rho} &\phantom{-}0.372 & 1.40 & \phantom{-} 0.437 \\
A_1^{B \rar \rho} &\phantom{-}0.261 & 0.29 & -0.415 \\ 
A_2^{B \rar \rho} &\phantom{-}0.223 & 0.93 & -0.092 \\
V^{B \rar \rho}   &\phantom{-}0.338 & 1.37 & \phantom{-} 0.315\\
T_1^{B \rar \rho} &\phantom{-}0.143 & 1.41 & \phantom{-} 0.361\\ 
T_2^{B \rar \rho} &\phantom{-}0.143 & 0.28 & -0.500 \\
T_3^{B \rar \rho} &\phantom{-}0.101 & 1.06 & -0.076 \\ \hline
\end{array}
$$   
\caption{The form factors for $B\rightarrow \rho \ell^{+}\ell^{-}$
in a three--parameter fit.}
\renewcommand{\arraystretch}{1}
\addtolength{\arraycolsep}{-3pt}
\end{table}

As it comes to the values of the Wilson coefficients $C_7^{eff}(m_b)$ and
$C_{10}(m_b)$, whose analytical expressions in the standard model
are given in \cite{R20,R21}, are
strictly real as can be read off from Table 2. 
In the leading logarithmic approximation, at the scale       
${\cal O}(\mu=m_b)$, we have                          
\bea                                                                         
C_7^{eff}(m_b) &=& -0.313~,\nnb \\             
C_{10}^{eff}(m_b) &=& -4.669~.\nnb                                           
\eea

\begin{table}[ht]
\renewcommand{\arraystretch}{1.5}
\addtolength{\arraycolsep}{3pt}
$$
\begin{array}{|c|c|c|c|c|c|c|c|c|c|}
\hline
C_{1} & C_{2} & C_{3} & C_{4} & C_{5} & C_{6}
& C_{7}^{eff} &
C_{9} & C_{10}^{eff} & C^{(0)} \\ \hline
-0.248 & 1.107 & 0.011 & -0.026 & 0.007 & -0.031 & -0.313 & 4.344 & -4.669
 & 0.362 \\ \hline  
\end{array}
$$
\caption{The numerical values of the Wilson coefficients at $\mu\sim m_{b}$
scale within the SM.}
\renewcommand{\arraystretch}{1}
\addtolength{\arraycolsep}{-3pt}
\end{table}
 
Although individual Wilson coefficients at $\mu \sim m_b$ level are all real
(see Table 2), the effective Wilson coefficient
$C_{9}^{eff}(m_{b},\hat s)$ has a finite phase, and in next--to--leading
order

\bea
\label{c9} 
C_9^{eff}(m_b,\hat s) = C_9(m_b)\left[1 + \frac{\alpha_s(\mu)}{\pi} \omega
(\hat s) \right]
+ Y_{SD}(m_b,\hat s) + Y_{LD}(m_B,s) ~,
\eea
where $C_9(m_b)=4.344$. Here $\omega \ga \hat s \dr$ represents the
${\cal{O}}(\alpha_{s})$ corrections
coming from one--gluon exchange in the matrix element of the corresponding
operator, whose explicit form can be found in \cite{R20}.
In (\ref{c9}) $Y_{SD}$ and $Y_{LD}$ represent, respectively, the short-- and
long--distance contributions of
the four--quark operators ${\cal{O}}_{i=1,\cdots,6}$ \cite{R20,R21}. Here
$Y_{SD}$ can be obtained
by a perturbative calculation  
\bea
Y_{SD}\ga m_{b}, \hat s \dr &=& g \ga \hat m_c,\hat s \dr
C^{(0)}
- \frac{1}{2} g \ga 1,\hat s \dr
\left[4 C_3 +4 C_4 + 3 C_5 + C_6 \right] \nnb \\
&-& \frac{1}{2} g \ga 0,\hat s \dr
\left[ C_3 + 3  C_4 \right]
+ \frac{2}{9} \left[ 3 C_3 + C_4 + 3 C_5 + C_6 \right] \nnb \\
&-& \lambda_u 
\left[ 3 C_1 + C_2 \right] \left[ g \ga 0,\hat s \dr -
g \ga \hat m_c,\hat s \dr \right]~,\nnb
\eea
where 
\bea
C^{(0)} &=& 3 C_1 + C_2 + 3 C_3 + C_4 + 3 C_5 + C_6~,\nnb \\ \nnb \\
\lambda_u &=& \frac{V_{ub} V_{ud}^\ast}{V_{tb} V_{td}^\ast}~, \nnb
\eea
and the loop function $g \ga m_q, s \dr$ stands for the loops
of quarks with mass $m_{q}$ at the dilepton invariant mass $s$.
This function develops absorptive parts for dilepton energies  
$s= 4 m_q^{2}$:
\bea
\lefteqn{
g \ga \hat m_q,\hat s \dr = - \frac{8}{9} \ln \hat m_q +
\frac{8}{27} + \frac{4}{9} y_q -
\frac{2}{9} \ga 2 + y_q \dr \sqrt{\vel 1 - y_q \ver}} \nnb \\
&&\times \Bigg[ \Theta(1 - y_q)
\ga \ln \frac{1  + \sqrt{1 - y_q}}{1  -  \sqrt{1 - y_q}} - i \pi \dr
+ \Theta(y_q - 1) \, 2 \, \arctan \frac{1}{\sqrt{y_q - 1}} \Bigg], \nnb
\eea
where  $\hat m_q= m_{q}/m_{b}$ and $y_q=4 \hat m_q^2/\hat s$. 
In addition to these perturbative contributions $\bar{c}c$ loops
can excite low--lying charmonium states $\psi(1s), \cdots, \psi(6s)$ 
whose contributions are represented by $Y_{LD}$ \cite{R22}.
However in the present work we restrict ourselves to the consideration of
short distance contributions only and therefore we find it redundant to
display the explicit form of $Y_{LD}$.

Calculations show that the asymmetry
${\cal A}_\varphi$ for the $B^0(t)\rar \pi^+ \pi^- \mu^+ \mu^-$ decay is 
sensitive only to tensor type interaction. We observe that, practically,
the existence of the new vector or scalar type interactions do not effect 
the SM results crucially, which is about $13\%$. However when tensor type
interactions are taken into consideration this asymmetry is reduced
noticeably and $\ga {\cal A}_\varphi \dr_{max} \sim 2.5\%$. Therefore if 
in future experiments the results obtained results differ from the SM
results, it is an unambiguous indication of the existence of new physics
beyond SM.

As one considers the $B^0(t)\rar \pi^+ \pi^- \tau^+ \tau^-$ decay, the
maximum value of the asymmetry ${\cal A}_\varphi$ is about $5\%$. 
Similarly, as in the $B^0(t)\rar \pi^+ \pi^- \mu^+ \mu^-$ decay, ${\cal
A}_\varphi$ is not sensitive to the new vector and scalar interactions and
for the contribution of the tensor interaction is less $1\%$.       

We further studied the dependence of 
the asymmetry ${\cal A}_1$ on the new Wilson coefficients and time for both
$B^0(t)\rar \pi^+ \pi^- \mu^+ \mu^-$ and $B^0(t)\rar \pi^+ \pi^- \tau^+
\tau^-$ decays. Our calculations show that the maximal possible value that
${\cal A}_1$ asymmetry gets for the $B^0(t)\rar \pi^+ \pi^- \mu^+ \mu^-$
decay is about $2\%$ for all type of new interactions. Similar behavior
takes place for the $B^0(t)\rar \pi^+ \pi^- \tau^+ \tau^-$ case as well,
with only one exception. For tensor type interactions which are controlled
by the new Wilson coefficients $C_T$ and $C_{TE}$,  ${\cal A}_1$ can
arrive at large values about $20\%$ (see Figs. (1) and (2)). So any departure 
from the SM prediction in future experiments, is an indication of the
presence of tensor type interactions.

Finally we have studied the dependence of the ${\cal A}_2$ asymmetry on new
Wilson coefficients and time for the above mentioned decays. For the 
$B^0(t)\rar \pi^+ \pi^- \mu^+ \mu^-$ decay the maximal value of ${\cal A}_2$
is about 
coefficient $C_{LL}=-4$ causes ${\cal A}_2$ to decrease about $50\%$ from
the SM prediction, but if $C_{LL}=+4$, practically the value of ${\cal A}_2$  
seems to coincide with the SM prediction, which also gives us a clue in
determining the sign of of the new vector interaction, as can be seen in
Fig. (3). As is depicted in Fig. (4), similar situation seems to hold 
for the new Wilson coefficient $C_{RL}$. Our numerical
analysis shows that ${\cal A}_2$ is quite insensitive to the existence of
scalar interaction and yields almost same results with SM prediction. 
But an important observation is that the tensor interaction enhances the 
SM prediction by about $50\%$. Therefore it would be an unambiguous
confirmation of the existence of new vector (tensor) interaction, if in
future experiments the measured value of ${\cal A}_2$  appears to be less
(larger) than the SM prediction.

On the other hand  for all values of the new Wilson coefficients,  
the ${\cal A}_2$ asymmetry for the $B^0(t)\rar \pi^+
\pi^- \tau^+ \tau^-$ case coincides almost with the SM prediction.
For this reason ${\cal A}_2$ for $B^0(t)\rar \pi^+ \pi^- \tau^+ \tau^-$   
decay does not seem to provide us with any new information about new physics
beyond SM.

As a final remark we should emphasize that similar analysis can be
performed for the $\bar B^0(t) \rar \pi^+ \pi^- \ell^+ \ell^-$ decay
as well. For this aim it is enough to use Eqs. (\ref{had}), (\ref{trn}) and
(\ref{trnn}).

In conclusion, in this work we have studied the time evolution of the decay
spectrum for the $B^0(t) \rar \pi^+ \pi^- \ell^+ \ell^-$ decay.
The sensitivity of the experimentally measurable asymmetries 
${\cal A}_\varphi$, ${\cal A}_1$ and ${\cal A}_2$ to the new Wilson
coefficients and time is studied in detail. It is observed that different
asymmetries show different dependencies on different new Wilson coefficients 
for the  $B^0(t)\rar \pi^+ \pi^- \mu^+ \mu^-$ and 
$B^0(t)\rar \pi^+ \pi^- \tau^+ \tau^-$ channels. Studying different
asymmetries for the above--mentioned two channels on new Wilson coefficients
can give essential information about new physics and can serve as an efficient
tool in determining not only of the magnitude, but also of the sign of 
new Wilson coefficients.

\newpage
\section*{Figure captions}
{\bf Fig. (1)} The dependence of the asymmetry ${\cal A}_1$ on the new
Wilson coefficient $C_T$ and time for $B^0(t)\rar \pi^+ \pi^- \tau^+ \tau^-$
decay, where $\tau_s$ is the $B^0$ meson life time.\\ \\
{\bf Fig. (2)} The same as in Fig. (1), but for the
new Wilson coefficient $C_{TE}$. \\ \\
{\bf Fig. (3)} The dependence of the asymmetry ${\cal A}_2$ on the new
Wilson coefficient $C_{LL}$ and time for $B^0(t)\rar \pi^+ \pi^- \mu^+ \mu^-$
decay.\\ \\
{\bf Fig. (4)} The same as in Fig. (3), but for the
new Wilson coefficient $C_{RL}$. \\ \\

\newpage     

\begin{figure}  
\vskip 1.5 cm   
    \includegraphics{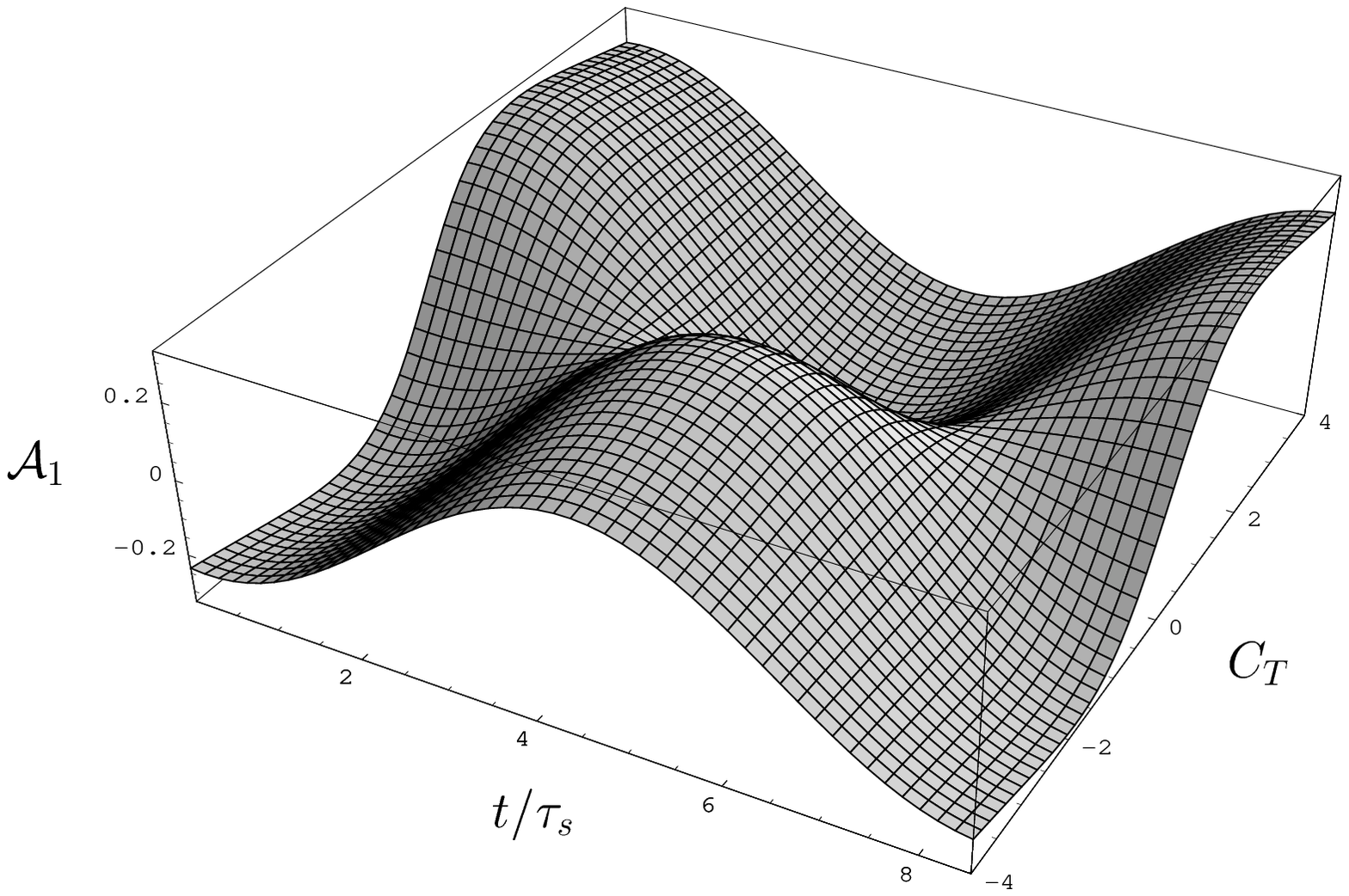}
\vskip 7.9cm     
\caption{}
\end{figure}

\begin{figure}
\vskip 1.5 cm
    \includegraphics{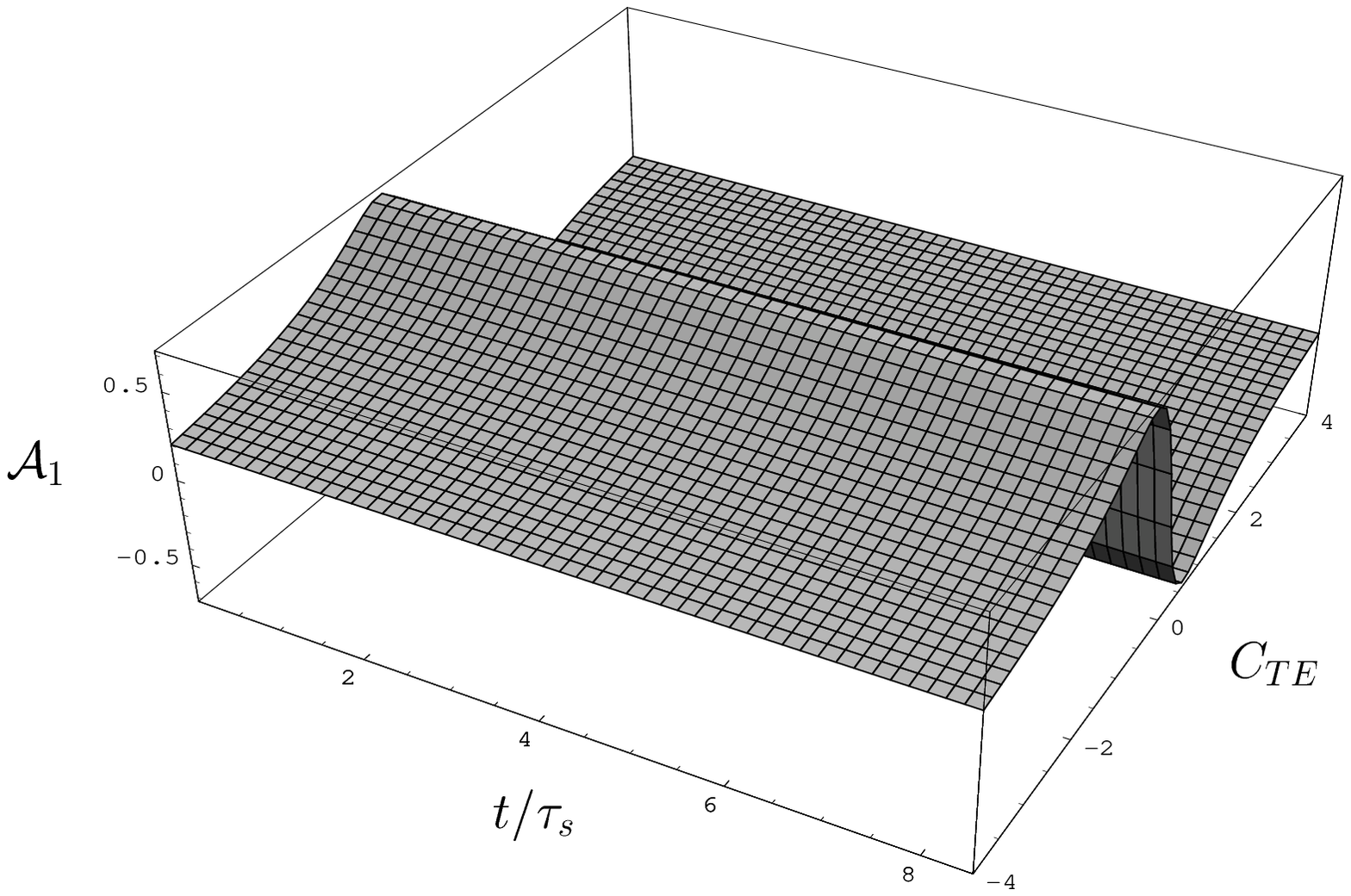}
\vskip 7.9 cm
\caption{}
\end{figure}

\begin{figure}  
\vskip 1.5 cm   
    \includegraphics{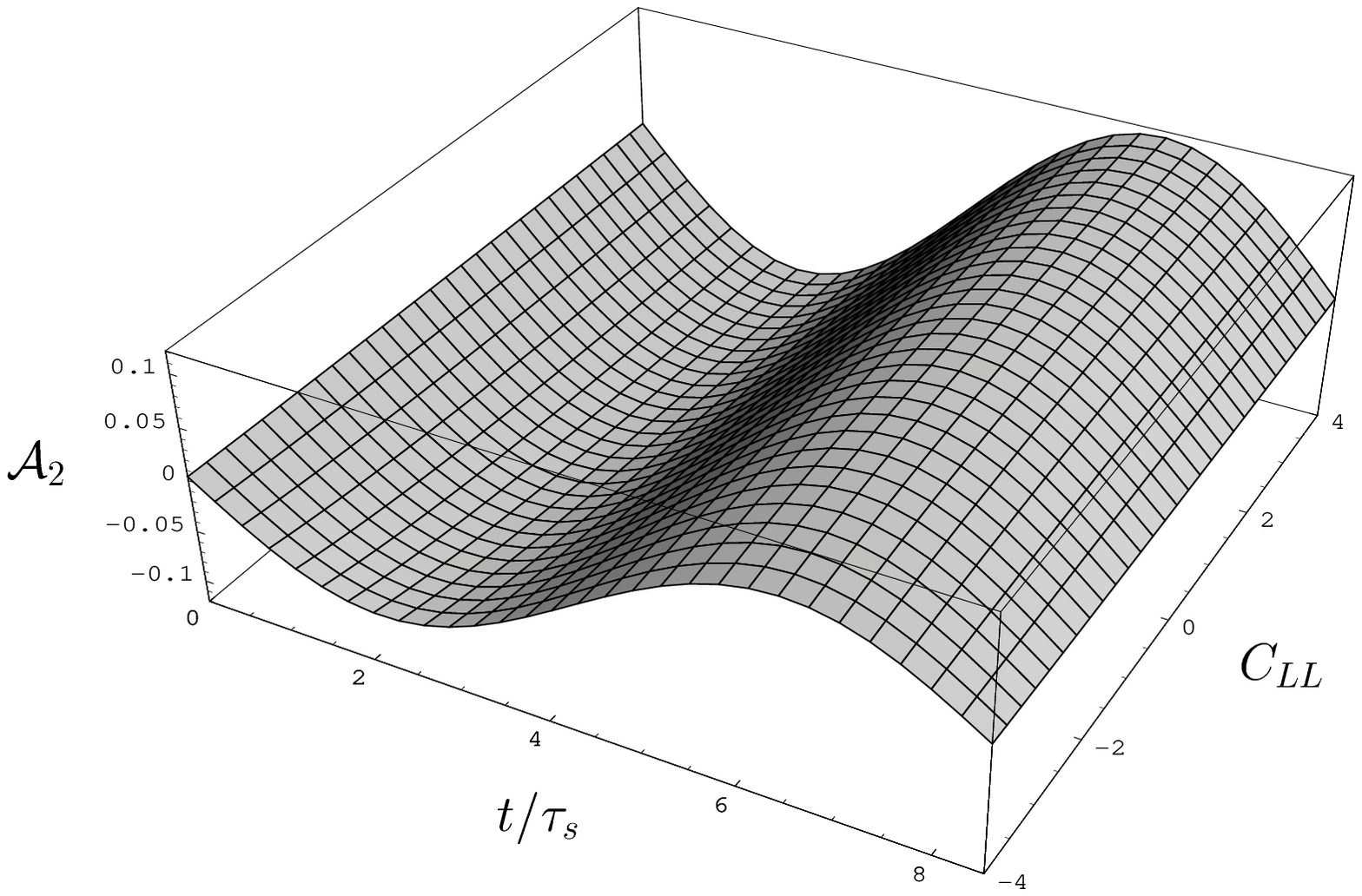}
\vskip 7.9cm     
\caption{}
\end{figure}

\begin{figure}
\vskip 1.5 cm
    \includegraphics{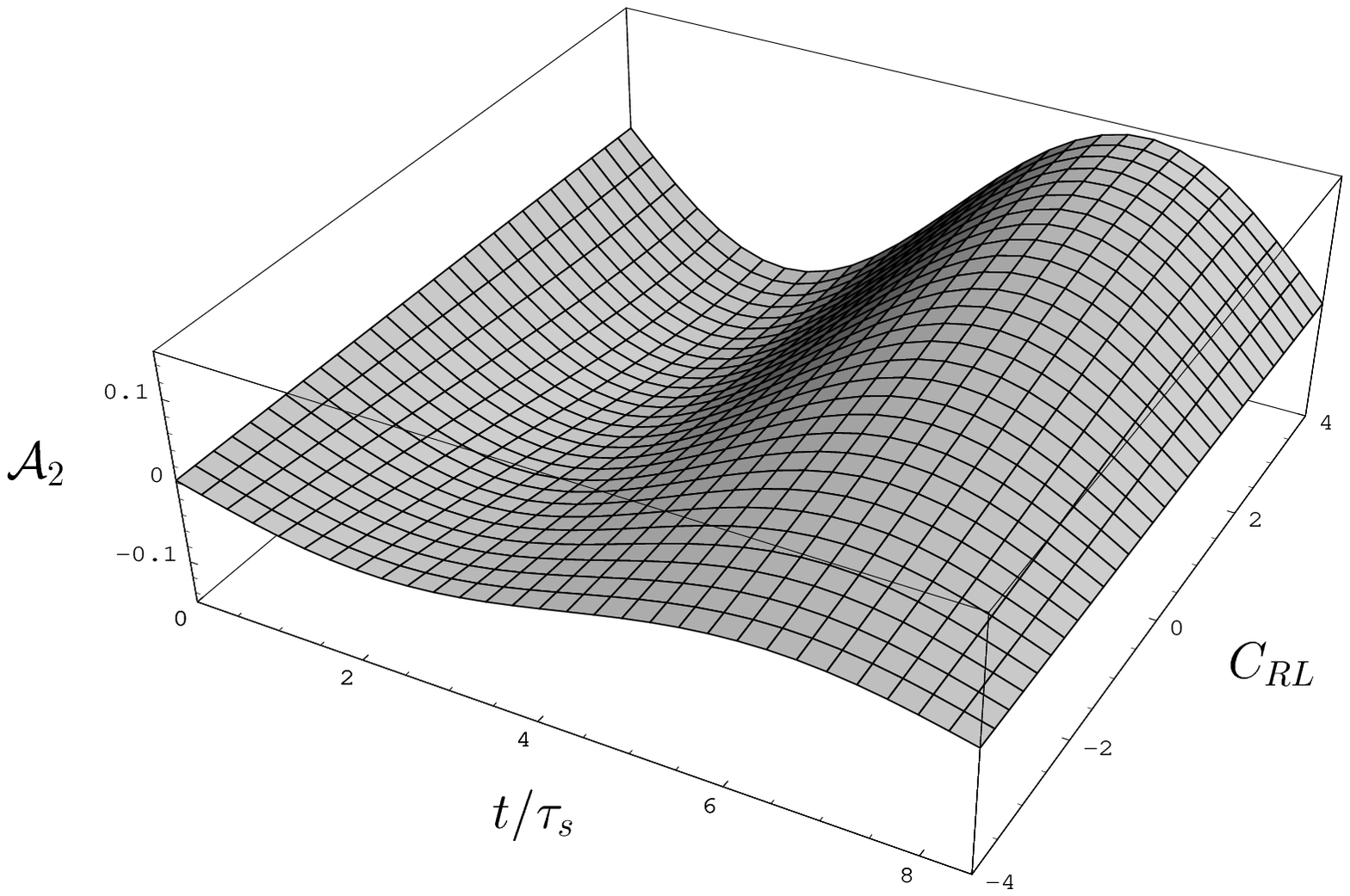}
\vskip 7.9 cm
\caption{}
\end{figure}


\newpage

\begin{thebibliography}{99}

\bibitem{R1} J. H. Christenson, J. W. Gronin, V. L. Fitch, R. Turlay,
{\it Phys. Rev. Lett.} {\bf D13} (1964) 138.
 
\bibitem{R2} KTeV Collaboration, A. Alavi--Harati {\it at al},
{\it Phys. Rev. Lett.} {\bf 84} (2000) 408.

\bibitem{R3} Y. Grossman, 
preprint: hep-ph/0012216 (2000).

\bibitem{R4} BaBar Collaboration, D. Hitlin,
preprint: hep-ex/0011024 (2000);\\
Belle Collaboration, H. Aihara,
preprint: hep-ex/0010008 (2000).

\bibitem{R5} G. Burdman and J. F. Donoghue,
{\it Phys. Rev.} {\bf D54} (1992) 187.

\bibitem{R6} NA48 Collaboration, A. Lai {\it at al},
{\it Phys. Lett.} {\bf B496} (2000) 137.

\bibitem{R7} L. M. Sehgal and J. van Leusen,
{\it Phys. Lett.} {\bf B489} (2000) 300.

\bibitem{R8} F. Kr\"{u}ger, L. M. Sehgal, N. Sinha and R. Sinha,
{\it Phys. Rev.} {\bf D61} (2000) 114028;\\
Erratum, {\it ibid} {\bf D63} (2001) 019901.

\bibitem{R9} T. M. Aliev and M. Savc{\i},
{\it Phys. Rev.} {\bf D62} (2000) 114010.
 
\bibitem{R10} S. Fukae, C. S. Kim and T. Yoshikawa,
{\it Phys. Rev.} {\bf D61} (2000) 074015.

\bibitem{R11} S. Fukae, C. S. Kim, T. Morozumi and T. Yoshikawa,
{\it Phys. Rev.} {\bf D59} (1999) 074013.

\bibitem{R12} T. M. Aliev, C. S. Kim and Y. G. Kim,,
{\it Phys. Rev.} {\bf D62} (2000) 014026.

\bibitem{R13} T. M. Aliev, D. A. Demir, M. Savc{\i},
{\it Phys. Rev.} {\bf D62} (2000) 074016.

\bibitem{R14} A. Pais and S. B. Treiman,
{\it Phys. Rev.} {\bf 168} (1968) 1858.

\bibitem{R15} G. Kramer and W. F. Palmer,
{\it Phys. Lett.} {\bf B279} (1992) 181.

\bibitem{R16} R. Sinha,
preprint: hep-ph/9608314 (1996).

\bibitem{R17} P. Ball and V. M. Braun,
{\it Phys. Rev.} {\bf D58}(1998) 094016.

\bibitem{R18}  P. Ball,
{\it JHEP} {\bf 9809} (1998) 005.

\bibitem{R19} T. M. Aliev, A. \"{O}zpineci and M. Savc{\i},
{\it Phys. Rev.} {\bf D55} (1997) 7059.

\bibitem{R20} M. Misiak,
{\it Nucl. Phys.} {\bf B393} (1993) 23; 
Erratum, {\it ibid} {\bf B439} (1995) 461.

\bibitem{R21} A. J. Buras and M. M\"{u}nz,
{\it Phys. Rev.} {\bf D52} (1995) 186.

\bibitem{R22} N. G. Deshpande, J. Trampetic and K. Panose,
{\it Phys. Rev.} {\bf D39} (1989) 1461;\\
C. S. Lim, T. Morozumi and A. I. Sanda,
{\it Phys. Lett.} {\bf B218} (1989) 343.

\end{thebibliography}
\end{document}